\documentclass[a4paper,12pt]{article}
\pdfoutput=1
\usepackage{graphicx,rotating}
\usepackage{hyperref}\usepackage{slashed}
\def\hhref#1{\href{http://arxiv.org/abs/#1}{#1}} 

\hypersetup{colorlinks,bookmarksopen,bookmarksnumbered,
linkcolor=blus,pdfstartview=FitH,urlcolor=rossos,citecolor=verdes}

\newcommand{\fig}[1]{~\ref{fig:#1}}
\usepackage{multicol}
\usepackage{color}
\definecolor{rosso}{cmyk}{0,1,1,0.4}
\definecolor{rossos}{cmyk}{0,1,1,0.55}
\definecolor{rossoc}{cmyk}{0,1,1,0.2}
\definecolor{blu}{cmyk}{1,1,0,0.3}
\definecolor{blus}{cmyk}{1,1,0,0.6}
\definecolor{bluc}{cmyk}{1,1,0,0.1}
\definecolor{verde}{cmyk}{0.92,0,0.59,0.25}
\definecolor{verdec}{cmyk}{0.92,0,0.59,0.15}
\definecolor{verdes}{cmyk}{0.92,0,0.59,0.4}

\newcommand{\eq}[1]{~{\rm (\ref{eq:#1})}}

\newcommand{\GeV}{\,{\rm GeV}}
\newcommand{\TeV}{\,{\rm TeV}}

\def\circa#1{\,\raise.3ex\hbox{$#1$\kern-.75em\lower1ex\hbox{$\sim$}}\,}

\newcommand{\NP}{Nucl. Phys.}

\newcommand{\PL}{Phys. Lett.}
\newcommand{\PR}{Phys. Rev.}

\newcommand{\beq}{\begin{equation}}
\newcommand{\eeq}{\end{equation}}

\font\tenrsfs=rsfs10 at 12pt
\font\sevenrsfs=rsfs7
\font\fiversfs=rsfs5
\newfam\rsfsfam
\textfont\rsfsfam=\tenrsfs
\scriptfont\rsfsfam=\sevenrsfs
\scriptscriptfont\rsfsfam=\fiversfs
\def\mathscr#1{{\fam\rsfsfam\relax#1}}

\def\circa#1{\,\raise.3ex\hbox{$#1$\kern-.75em\lower1ex\hbox{$\sim$}}\,}
\makeatletter

\def\art{\@ifnextchar[{\eart}{\oart}}
\def\eart[#1]#2#3#4#5#6{{\rm #2}, {#3 #4} {\rm (#6) #5} [arXiv:{\hhref{#1}}]}
\def\hepart[#1]#2{{\rm #2, arXiv:\hhref{#1}}}
\newcounter{alphaequation}[equation]
\def\thealphaequation{\theequation\hbox to
0.6em{\hfil\alph{alphaequation}\hfil}}
\def\eqnsystem#1{
\def\@eqnnum{{\rm (\thealphaequation)}}
\def\@@eqncr{\let\@tempa\relax \ifcase\@eqcnt \def\@tempa{& & &} \or
 \def\@tempa{& &}\or \def\@tempa{&}\fi\@tempa
 \if@eqnsw\@eqnnum\refstepcounter{alphaequation}\fi
\global\@eqnswtrue\global\@eqcnt=0\cr}
\refstepcounter{equation} \let\@currentlabel\theequation \def\@tempb{#1}
\ifx\@tempb\empty\else\label{#1}\fi
\refstepcounter{alphaequation}
\let\@currentlabel\thealphaequation
\global\@eqnswtrue\global\@eqcnt=0 \tabskip\@centering\let\\=\@eqncr
$$\halign to \displaywidth\bgroup \@eqnsel\hskip\@centering
$\displaystyle\tabskip\z@{##}$&\global\@eqcnt\@ne
\hskip2\arraycolsep\hfil${##}$\hfil& \global\@eqcnt\tw@\hskip2\arraycolsep
$\displaystyle\tabskip\z@{##}$\hfil
\tabskip\@centering&\llap{##}\tabskip\z@\cr}
\def\endeqnsystem{\@@eqncr\egroup$$\global\@ignoretrue} \makeatother

\newcommand{\eV}{\,{\rm eV}}

\newcommand{\SU}{\,{\rm SU}}
\begin{document}

\begin{center}
IFUP-TH/2008-12\bigskip\bigskip\bigskip\bigskip

{\Huge\bf\color{red}
Sommerfeld corrections to\\[5mm]
type-II and III leptogenesis}\\

\medskip
\bigskip\color{black}\vspace{0.6cm}
{
{\large\bf Alessandro Strumia}
}
\\[7mm]
{\it Dipartimento di Fisica dell'Universit{\`a} di Pisa and INFN, Italia}

\bigskip\bigskip\bigskip

{\large
\centerline{\large\bf Abstract}

\begin{quote}
We study thermal leptogenesis from decays of 
the electroweak triplets that mediate neutrino masses in type-II and type-III see-saw.
We find that Sommerfeld corrections reduce the baryon asymmetry by $\sim 30\%$,
and that successful leptogenesis needs triplets heavier than 1.6 TeV,
beyond the discovery reach of LHC.

\end{quote}}

\end{center}

\section{Introduction}
Leptogenesis might be produced by decays of scalar  (type-II see-saw) or fermion
(type-III see-saw~\cite{foot}) SU(2)$_L$ triplets with mass $M$.  The Boltzmann equation for their abundance contains their annihilation rate $\gamma_A$
(number of annihilations per spacetime volume at temperature $T$):
the final baryon asymmetry depends on the value of $\gamma_A$ in the non-relativistic limit, at $T\sim M/\ln(M_{\rm Pl}/M)$.
Non-relativistic scatterings among gauge-charged particles are affected by 
non-perturbative corrections: the states involved in the annihilation no longer are plane waves
when the kinetic energy is comparable to the electroweak potential energy~\cite{Sommerfeld}.
In this paper we study how such Sommerfeld corrections affect the final baryon asymmetry.

In section~\ref{S} we use group theory to derive a simple formula for non-abelian
Sommerfeld corrections.
In section~\ref{III} we apply it to type-III see-saw and in section~\ref{II} to type-II see-saw.

In both cases we also extend computations of thermal leptogenesis
down to values of $M\sim\TeV$ accessible to the LHC collider, motivated by
recent papers that studied how type-II or III see-saw can be searched for at LHC.
However, we find  a model-independent lower bound on $M$ that limits this possibility.
In section~\ref{concl} we discuss our results and conclude.

\section{Sommerfeld corrections}\label{S}
We denote as ${\cal T}$ the scalar or fermion $T$riplet that decays producing the lepton asymmetry.
Following~\cite{LeptogenesisTypeIII,HRS}
the final baryon asymmetry is parameterized as $n_B/n_\gamma=-0.029\eta\varepsilon$,
where the efficiency $\eta$ encodes the dynamics and $\varepsilon$ the amount of CP violation.
The Boltzmann equation that dictates the evolution of the triplet abundance
contains the ${\cal T}$
decay rate $\gamma_D$ and the ${\cal TT}^*$ annihilation rate
$\gamma_A$:
\beq
sHz \frac{dY}{dz} =
  -\bigg(\frac{Y}{Y^{\rm eq}}-1\bigg)\gamma_D
  -2\bigg(\frac{Y^2}{Y^{2\rm eq}}-1\bigg)\gamma_A \,. \eeq
where 
$z\equiv M/T$, 
 $H$ is the Hubble rate at temperature $T$;
$Y\equiv n_{\cal T}/s$ for the real fermion triplet 
and $Y\equiv (n_{\cal T} + n_{{\cal T}^*})/s$ for the complex scalar triplet;
  $Y_{\rm eq}$ is the value that $Y$ would have in thermal equilibrium;
$n$ is the number density; $s={2\pi^2}g_{*s}T^3/45$ with $g_{*s}=106.75$ is  the entropy density of SM particles;
$\gamma_A$ is the thermal average of the annihilation cross section 
  summed over all initial- and final-state Lorentz and gauge indices: 
   \beq 
  \gamma_A =\frac{T}{64 \pi^4} \int_{4M^2 }^{\infty} ds~ s^{1/2}
 {\rm K}_1\bigg(\frac{\sqrt{s}}{T}\bigg) \hat{\sigma}_A(s)\qquad\hbox{where}\qquad
 \hat\sigma_A\equiv\int dt \sum_{\rm all}\frac{|\mathscr{A}|^2}{8\pi s},\eeq
and K$_1$ is a Bessel function; 

Explicit expressions for $\hat\sigma_A$ are given in appendix A of~\cite{MDM2} for a generic $\SU(2)_L\otimes{\rm U}(1)_Y$ multiplet.
Similarly to the case of Cold Dark Matter freeze-out, the value of $\gamma_A$
is relevant in the non relativistic limit, $T/ M\approx 1/\ln M_{\rm Pl}/M \ll 1$, 
so that
$\hat\sigma_A$ can be approximated as the sum of $s$-wave, $p$-wave, etc contributions:
\beq \hat\sigma_A = c_s \beta + c_p \beta^3 +{\cal O}(\beta^5)\eeq
where $\beta = \sqrt{1-4M^2/s}$ is the triplet velocity in the ${\cal TT}^*$  center of mass frame.
The corresponding annihilation rate is
\beq
 \gamma_A=
\frac{MT^3e^{-2M/T}}{32\pi^3}\left[c_s + \frac{3T}{2M}(c_p+\frac{c_s}{2})+{\cal O}(\frac{T}{M})^2\right].
\eeq
For the fermion triplet one has 
\beq \label{eq:III} c_s=\frac{111g_2^4}{8\pi},\qquad c_p = \frac{51 g_2^4}{8\pi}.\eeq
For the scalar triplet $\gamma_A$ is well approximate by just its $s$-wave coefficient
\beq \label{eq:II}c_s=\frac{9g_2^4+12 g_2^2 g_Y^2 + 3g_Y^4}{2\pi}
\eeq
(hypercharge is normalized such that $Y=1/2$ over  lepton doublets).

\bigskip

We now compute the non-perturbative electroweak Sommerfeld corrections to $c_s$.
Scatterings among charged particles 
are distorted by the Coulomb force, when their kinetic energy is
low enough that the electrostatic potential energy is relevant.
This leads e.g.\ to significant enhancements of the $\mu^-\mu^+$
annihilation cross section (attractive force) or to significant
suppressions of various nuclear processes (repulsive force).
The Sommerfeld correction to $s$-wave
point-like annihilations (e.g.\ $\mu^-\mu^+\to\gamma\gamma$) 
can be computed as~\cite{Sommerfeld}
$S=|\psi(\infty)/\psi(0)|^2$, where $\psi(r)$ is the (reduced) $s$-wave-function for the two-body
state with kinetic energy $K=M\beta^2$, that in the non-relativistic limit 
satisfies
the  Schr\"odinger equation
\beq \label{eq:S}
-\frac{1}{M}\frac{d^2 \psi}{dr^2}+ V \cdot \psi= K \psi
\eeq
with outgoing boundary condition $\psi'(\infty)/\psi(\infty) \simeq i M\beta$.
For a single abelian massless vector with potential $V=\alpha/r$ the result is
$\sigma = S \sigma_{\rm perturbative}$ where the Sommerfeld correction is~\cite{Sommerfeld} 
\beq  \label{eq:R0}
S(x) = \frac{-\pi x}{1-e^{\pi x}} \qquad x = \frac{\alpha}{\beta}.\eeq
Here $\alpha<0$ describes an attractive potential that leads to an enhancement $S> 1$, and
$\alpha>0$ describes a repulsive potential that leads to $S<1$.
We will need the thermally averaged value of the Sommerfeld correction:
\beq S_T(y) = \frac{\int_0 \beta^2 e^{-M\beta^2/T} S(\alpha/\beta)~d\beta}{\int_0 \beta^2 e^{-M\beta^2/T} ~d\beta},\qquad
y\equiv \frac{\alpha}{\beta_T}\eeq
where $\beta_T\equiv \sqrt{T/M}$ is the characteristic velocity in the thermal bath
and the upper integration limit on $\beta$ is any value much larger that $\beta_T$.
The function $S_T$ is computed numerically, with the qualitative result $S_T(x)\approx S(x)$.

\bigskip

The generalization to non-abelian massive vectors was discussed in~\cite{hisano,MDM2} and
needs a long list of potential and annihilation matrices, 
such that the matrix Schr\"odinger equation\eq{S} can be solved only numerically.
We here show how in the SU(2)$_L$-invariant limit $M\gg M_{W,Z}$ the Sommerfeld correction can be analytically computed
as a sum of abelian-like cases.
For any simple gauge group with coupling $\alpha=g^2/4\pi$, the potential between
two particles in representations $R$ and $R'$ at distance $r$ is
\beq V=\frac{\alpha}{r}\sum_a T^a_{R}\otimes T^a_{R'}\eeq
where $a$ runs over the adjoint,
$T^a_R$ are the generators in the representation $R$, and
the tensor product $\otimes$ indicates that $V$ has 4 gauge indices.
Unlike in~\cite{MDM2} we do not include (anti)symmetrizations and their corresponding normalization of the 2-body states:
their only effect is enforcing $V=0$ for the 2-fermion states forbidden by the Pauli exclusion principle,
which can be less formally imposed by hand.
Group theory allows to decompose the tensor product of the representations $R$ and $R'$
into a sum of irreducible representations $Q$:
\beq \label{eq:TAxB} R\otimes R'=\bigoplus_Q Q,\qquad
T^a_{R\otimes R'} \equiv T^a_R \otimes 1_{R'} +1_R\otimes T^a_{R'} =\sum_Q T^a_Q\eeq
so that, recalling the definition $T^a_R \cdot T^a_R\equiv C_R 1_R$ of the quadratic Casimir $C_R$, gives:
\beq V=\frac{\alpha}{2r}\sum_A (T^a_{R\otimes R'}\cdot T^a_{R\otimes R'} - T^a\cdot T^a\otimes 1_{R'} - 1_R\otimes T^A_{R'}\cdot T^A_{R'})=
\frac{\alpha}{2r} (\sum_Q C_Q 1_Q -C_R  1 -C_{R'}1)
\eeq
where $1_Q$ is the projector along the subspace $Q$, so that the matrix
$V$ is diagonal along each irreducible representation $Q$.
The Casimir of the SU(2) irreducible representations with dimension $n$ is $C_n=(n^2-1)/4$,
so that a two-body state $n\otimes\bar n$ with total isospin $N\le 2n-1$ has potential
$V = (N^2+1-2n^2)\alpha_2/8r$.
The two-body state of the (scalar or fermion)
triplets involved in leptogenesis decompose as $3\otimes 3 = 1_S \oplus 3_A\oplus 5_S$,
and the potentials are $V=-2\alpha_2/r$ for the singlet state,
$V=-\alpha_2/r$ for the triplet state, and
$V=\alpha_2/r$ for the quintuplet state.

\bigskip

Finally, one might worry that the Sommerfeld correction, computed for scattering in vacuum,
does not apply for annihilations in the cosmological plasma.
This is not the case because $\gamma_A$ is only needed when the ${\cal TT}^*$ annihilation rate
becomes smaller than the expansion rate $H\sim T^2/M_{\rm Pl}$:
in view of the Planck suppression $H$ and thereby $\gamma_A$ is so small that annihilations are rare
enough that the vacuum approximation holds.
Furthermore, the ${\cal T}$ life-time $\tau \sim 1/(\lambda^2 M)$ is longer enough than
the Coulomb time-scale $\sim 1/(M\beta^2)$ provided that $\lambda \ll \beta$,
where $\lambda$ denotes the small coupling(s) present in type-III or type-II see-saw.
Finally, thermal effects generate a Debye mass for the vectors
($m^2 = {11}g_2^2T^2/6$ for the $\SU(2)_L$ vectors), 
screening the Coulomb potential into a Yukawa potential,
$e^{-mr}/r$: the results of~\cite{MDM2} show that the Debye mass
$m$ is small enough that we can neglect it, approximating vectors as massless.

\begin{figure}[t]
\begin{center}
$$\includegraphics[width=0.45\textwidth]{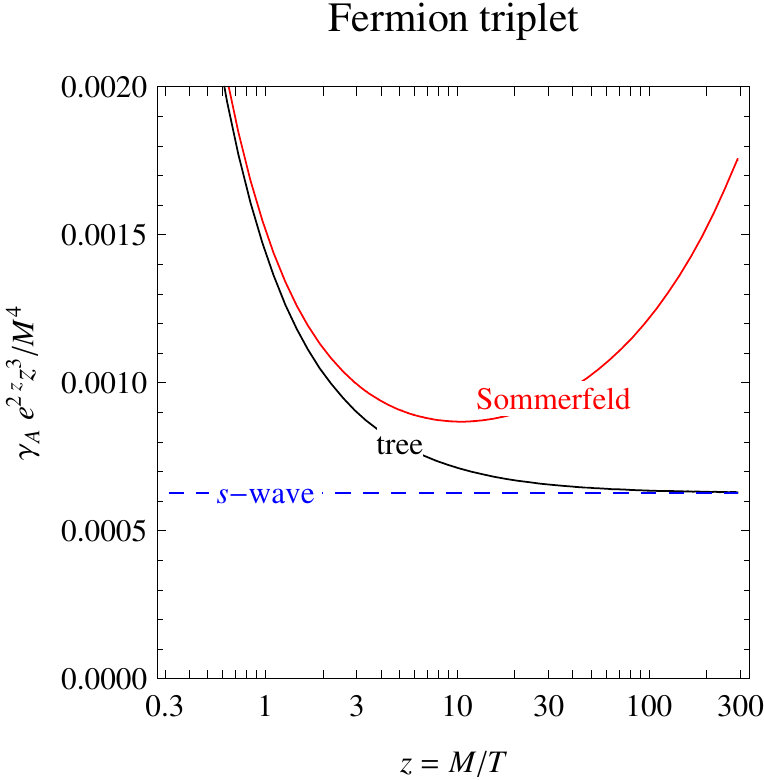}\qquad
\includegraphics[width=0.45\textwidth]{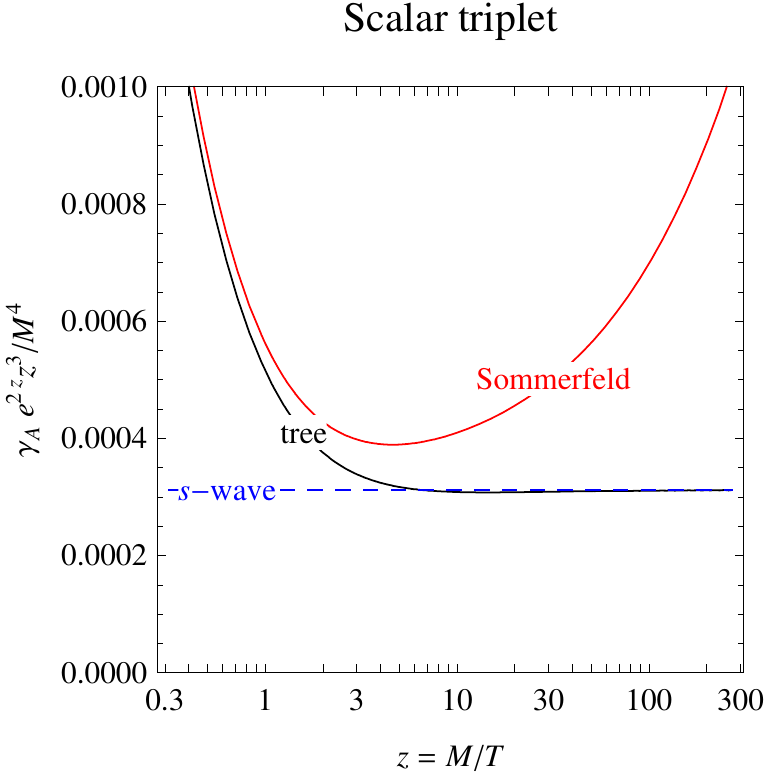}$$
\caption{\em Annihilation rate $\gamma_A$ for the fermion and scalar triplet for $M=10^5\GeV$.
The `tree' line shows the tree-level result, the dashed line shows the $s$-wave approximation, and the `Sommerfeld'
line includes Sommerfeld corrections.  Leptogenesis is dominantly produced at $z\equiv M/T \sim 20$.
\label{fig:ann}}
\end{center}
\end{figure}

\begin{figure}[t]
\begin{center}
$$\includegraphics[width=0.45\textwidth]{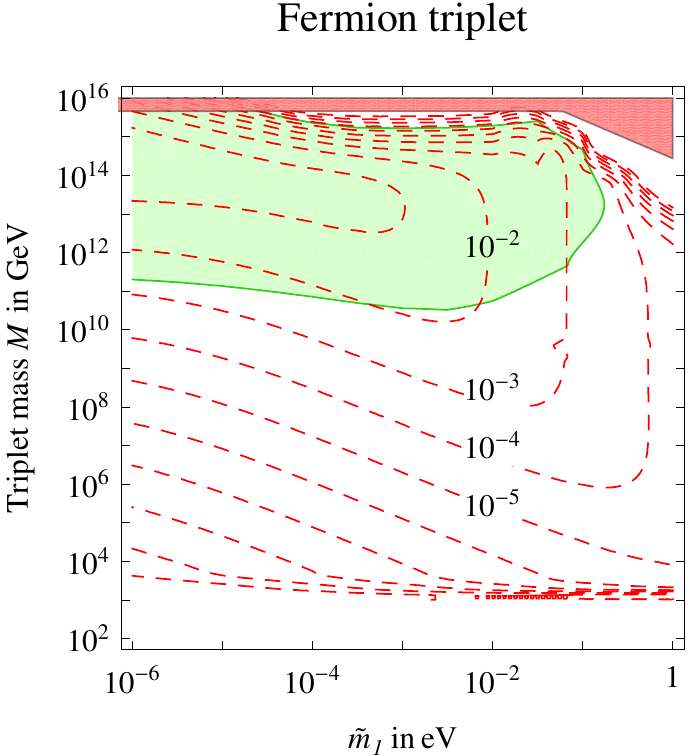}\qquad
\includegraphics[width=0.45\textwidth]{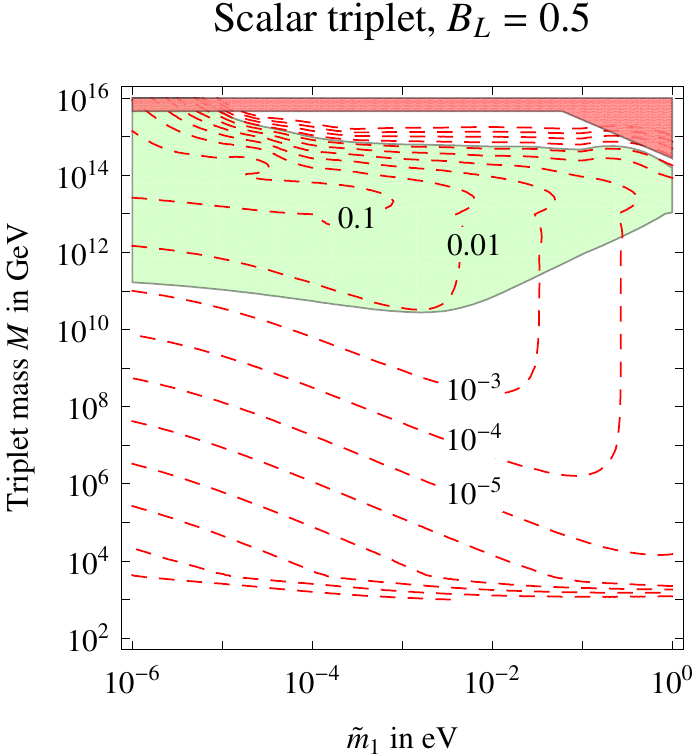}$$
\caption{\em Iso-contours of the efficiency $\eta$ of thermal leptogenesis from decays of a fermion triplet (left, type-III see-saw) and a scalar triplet (right, type-II see-saw, in the case of equal branching ratio into leptons and into higgses).
The regions shaded in green are allowed by the model-dependent bound on the CP asymmetry generated by the neutrino mass operator $(LH)^2$.
\label{fig:eta}}
\end{center}
\end{figure}

\begin{figure}[t]
\begin{center}
$$\includegraphics[width=0.45\textwidth]{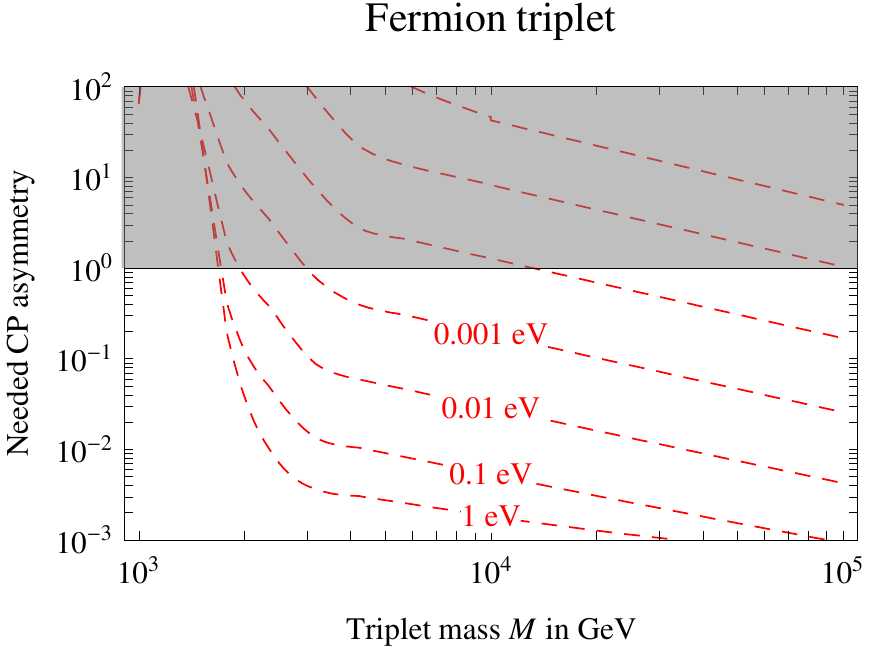}\qquad
\includegraphics[width=0.45\textwidth]{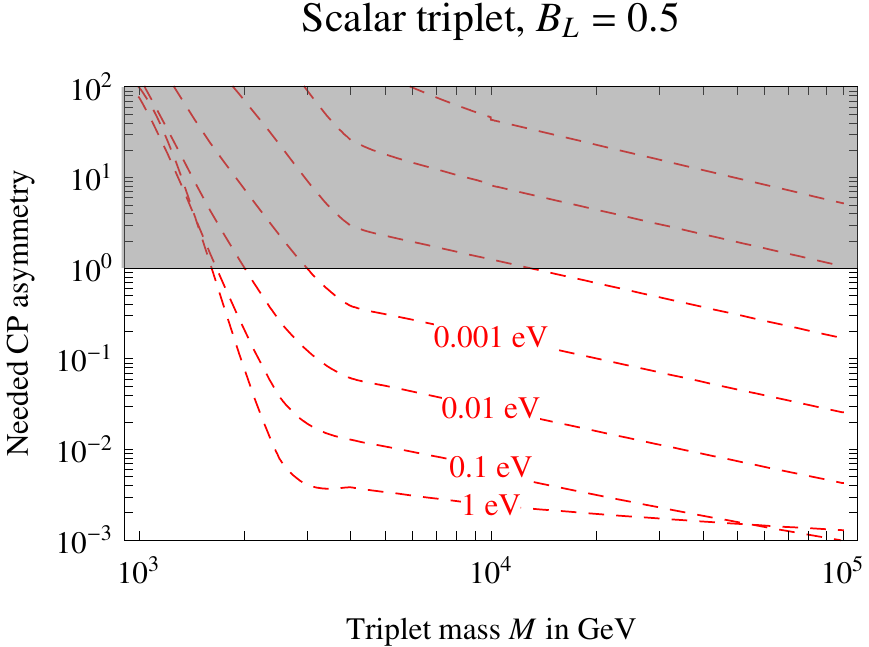}$$
\caption{\em Values of the {\rm CP} asymmetry $|\varepsilon_1|\le 1$ 
needed to get the observed
baryon asymmetry from decays of a fermion triplet (left) and a scalar triplet (right),
as function of its mass $M$ for the values of  $\tilde{m}_1$ superimposed to the curves.
We see that successful leptogenesis needs $M>1.6\TeV$.
\label{fig:eta2}}
\end{center}
\end{figure}

\section{Type-III see-saw: leptogenesis from a fermion triplet}  \label{III}
The two body ${\cal TT}$ states can be classified
according to their quantum numbers ($I,S,L$), where $I=\{1,3,5\}$ is the total isospin, $S=\{0,1\}$ is the total spin,
$L$ is total orbital angular momentum. 
Restricting to the dominant $s$-wave annihilations, $L=0$,
the states allowed by quantum statistics are
$(I,S)=(1,0)$, $(3,1)$ and $(5,0)$.
Their annihilations into couples of SM vectors ($W^a$ and $Y$ of $\SU(2)_L$ and hypercharge),
of SM fermions $\Psi$ and of SM Higgs doublets $H$ are restricted by their $(I,S,L)$ quantum numbers.
Taking into account the potentials computed in section~\ref{S},
 the total Sommerfeld-corrected $s$-wave annihilation rate is:\footnote{We compare with previous results.
 Neglecting Sommerfeld corrections, $S_T\to 1$:
 i) eq.\eq{SF} agrees with eq.~(29c) of~\cite{LeptogenesisTypeIII} 
 (where a factor 2 was however missed in the Boltzmann equation in front of $\gamma_A$;
 the factor 1/24 coming from annihilations into $HH^*$ was neglected;
the $g_2^4$ factor was unproperly typed as $g_2^8$ in the numerical code);
ii) eq.\eq{SF} differs by order one factors from eq.~(4.9) of~\cite{LeptogenesisTypeIIIb};
iii) eq.\eq{SF} agrees with analogous computations of the relic abundance of wino-like dark matter.
Some dark matter studies also included Sommerfeld corrections~\cite{hisano,MDM2},
but a direct comparison is not immediate (dark matter studies included $\SU(2)_L$-breaking effects 
and thereby used electric charge instead of isospin to classify states, obtaining complex expressions);
the numerical results for $\gamma_A$ can be compared in the limit $M\gg M_Z/\alpha_2$ and agree.
Some recent experimental results suggested the possibility of
Sommerfeld enhancememnts from new gauge interactions in the Dark Matter sector~\cite{DMSom}:
our result in eq.\eq{TAxB} holds for any gauge group and allows to study such effects.}
\beq \label{eq:SF} c_s =   
\underbrace{\frac{2 g_2^4}{\pi} S_T(-2\alpha_2 \sqrt{z})}_{(I,S)=(1,0)\to W^a W^a} + 
\underbrace{\frac{5g_2^4}{2\pi} S_T(\alpha_2 \sqrt{z})}_{(I,S)=(5,0)\to W^a W^b} +
\underbrace{(1+\frac{1}{24})\frac{9g_2^4}{\pi}S_T(-\alpha_2\sqrt{z})}_{(I,S)=(3,1)\to \Psi\Psi,HH}
 \eeq
 where the text under each contribution to the rate indicates the corresponding annihilation processes.
 Eq.\eq{SF} differs from the perturbative result of eq.\eq{III} by an order one factor
 at $T\circa{<} M g_2^2$.
 
Fig.\fig{ann}a shows our result for the adimensional combination $\gamma_A e^{2z}z^3/M^4$, that becomes constant
in the non-relativistic limit $z\equiv M/T\gg1$, where $s$-wave annihilations dominate.
The final baryon asymmetry roughly depends on  $\gamma_A$ at $z\sim 20$:
the Sommerfeld  enhancement to the $s$-wave contribution is
more important than including $p$-wave and all other $L\neq 0$ annihilations.

Fig.\fig{eta}a shows the contour plot of the efficiency $\eta$ as function of $M$ and of $\tilde{m}_1$,
which, as usual, is the lightest triplet contribution to neutrino masses.
Considering e.g.\ the point $M=10^5\GeV$ and $\tilde{m}_1  =10^{-5}\eV$, Sommerfeld corrections
reduce the efficiency from $2.6~10^{-8}$ to $2.0~10^{-8}$.

For large enough $\tilde{m}_1$
the decay rate $\gamma_D$ is larger than the annihilation rate $\gamma_A$ in the relevant temperature range $T\sim M/20$,
so that the efficiency $\eta$ no longer depends on $\gamma_A$,
and gets the value typical of type-I see-saw, largely independent on $M$.
In this region Sommerfeld corrections, that enhance  $\gamma_A$,
do not affect the final asymmetry.

As in~\cite{LeptogenesisTypeIII} we shaded in green the region where thermal leptogenesis can 
produce the observed baryon asymmetry, assuming that the CP asymmetry is generated
by two other triplets so heavy that their effects are fully encoded in the dimension-5 operator $(LH)^2$,
constrained by neutrino masses.
We see that the resulting bound on the CP asymmetry~\cite{LeptogenesisTypeIII}
implies  the model-dependent lower bound $M>3~10^{10}\GeV$.

\bigskip

Various recent papers considered type-III and type-II see-saw with $M$ so small that triplets  can be probed by colliders:
LHC experiments are going to probe the region $100\GeV\circa{<}M\circa{<}{1}\TeV$.
We here find that this range of $M$ is not compatible with thermal leptogenesis,
that needs $M>1.6\TeV$ because at lower $M$ the efficiency becomes so small that the baryon asymmetry $n_B/n_\gamma = -0.029 \eta\varepsilon_1$~\cite{LeptogenesisTypeIII}
remains smaller than the observed value, $n_B/n_\gamma \approx 6.15~10^{-10}$~\cite{review},
even if the CP asymmetry is maximal, $|\varepsilon_1|=1$ (resonant leptogenesis~\cite{Pilaftsis}
allows to realize an order one CP-asymmetry,
although unity is not reached).
This issue is better illustrated by fig.\fig{eta2}a, that also allows to see how the lower bound on $M$ depends on $\tilde{m}_1$.

\begin{figure}[t]
\begin{center}
$$\includegraphics[width=0.45\textwidth]{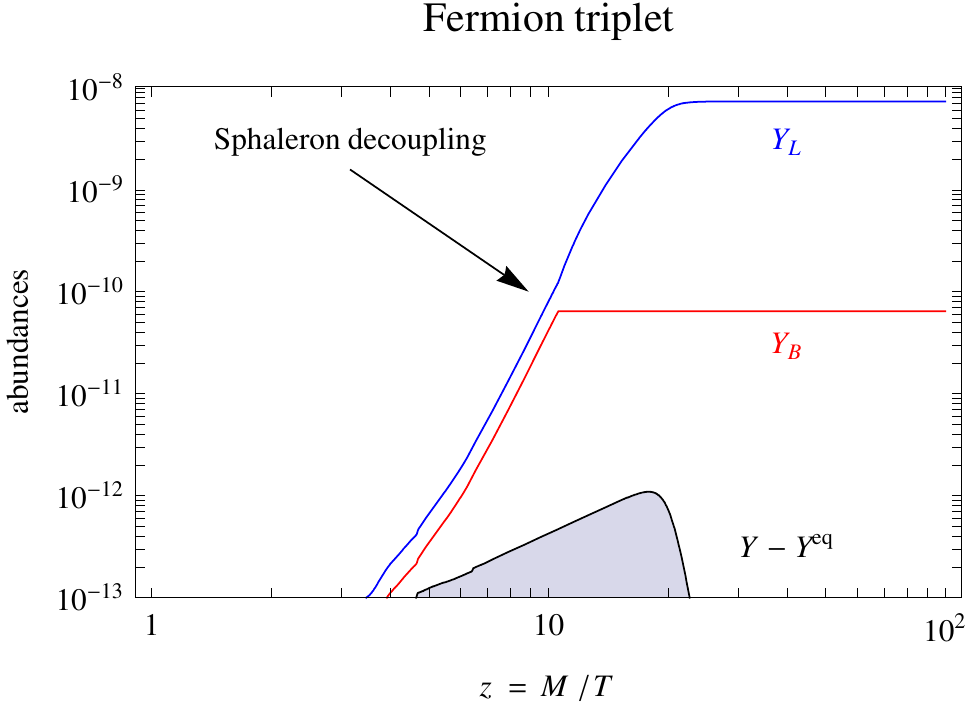}\qquad
\includegraphics[width=0.45\textwidth]{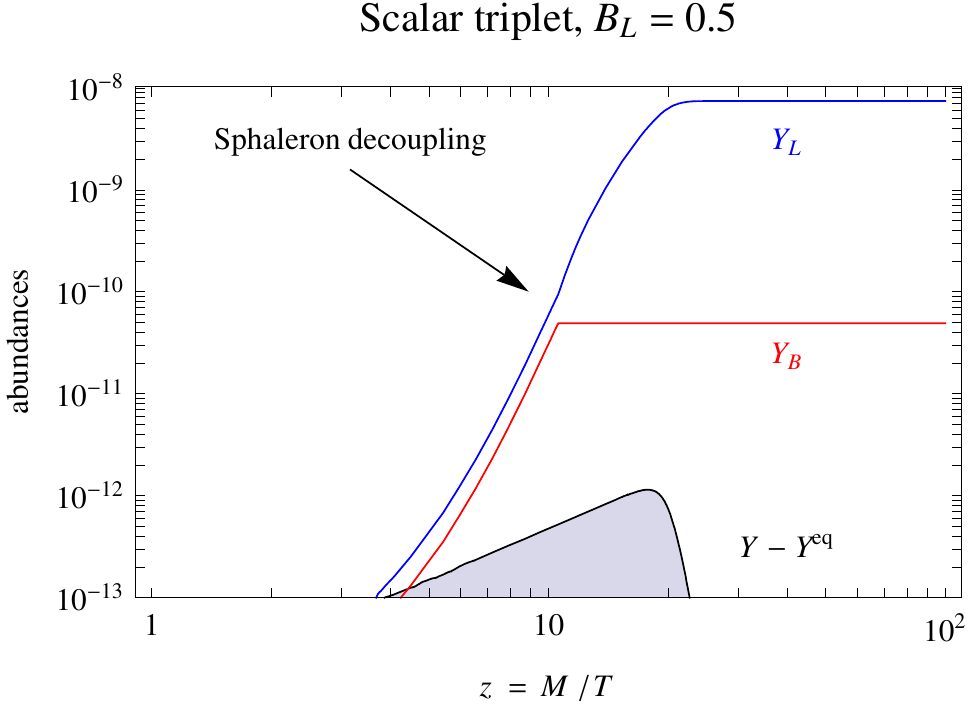}$$
\caption{\em Evolution of $Y_L$, $Y_B$, $Y-Y^{\rm eq}$ for $M=1.6\TeV$,
$\tilde{m}_1 = m_{\rm atm} = 0.06\eV$, $|\varepsilon_1| = 1$.
\label{fig:sample}}
\end{center}
\end{figure}

Let us now discuss why the efficiency strongly decreases when $M$ is below 3 TeV.
The Higgs starts to acquire its vev via a first order phase transition at the critical temperature $T_{\rm cr}\approx 1.2 m_h$.
Sommerfeld corrections have to be computed as in~\cite{hisano,MDM2}, but this is a minor detail.
The key issue is that $\SU(2)_L$-breaking suppresses the sphaleron rate $\gamma_{\rm sphaleron}$~\cite{Shap},
so that the lepton asymmetry ceases to be converted into the baryon asymmetry below
the temperature $T_{\rm dec}\approx 80\GeV+0.45 m_h$~\cite{Shap}.
The approximations for $T_{\rm cr}$ and $T_{\rm dec}$ hold for the light Higgs mass suggested by present data, 
$115\GeV\circa{<}m_h\circa{<}200\GeV$~\cite{LEP}; 
we here assume $m_h=120\GeV$.
In order to quantitatively include the sphaleron decoupling effect it is convenient to write
the Boltzmann equations for the ${\cal B}-{\cal L}$ and the $\cal B$ asymmetries:
\begin{eqnsystem}{sys:BoltzBL}
sHz \frac{dY_{{\cal B} - {\cal L}}}{dz} &=&
-\gamma_D \varepsilon_{1} \bigg(\frac{Y}{Y^{\rm eq}}-1\bigg)  
-\frac{Y_{{\cal B} - {\cal L}}}{Y_{L}^{\rm eq}}\frac{\gamma_D}{2} \\
sHz \frac{dY_{{\cal B}} }{dz} &=& -\gamma_{\rm sphaleron} ( Y_{\cal B} c - (1+c) Y_{{\cal B}-{\cal L}})  
\qquad c\approx0.52
\label{eq:BoltzB}
\end{eqnsystem}
The first equation  is the standard one~\cite{LeptogenesisTypeIII}, as sphalerons
conserve ${\cal B}-{\cal L}$.\footnote{Including flavor details the first equation gets replaced by
three equations for the partial ${\cal B}/3-{\cal L}_{e,\mu,\tau}$ asymmetries~\cite{BCST}.
If fermion triplet(s) will be discovered at LHC, it should be possible to reconstruct the flavor structure of their couplings~\cite{LHCtypeIII}
and a more precise analysis will become worthwhile.

}
The second equation tells the final baryon asymmetry, and is approximatively solved by 
\beq Y_B(T<T_{\rm dec}) \approx (1+1/c)Y_{{\cal B}-{\cal L}}(T_{\rm dec}).\eeq
The evolution of the various asymmetries is illustrated in fig.\fig{sample}a, where we see that the observed baryon asymmetry is reproduced with a maximal
CP asymmetry $|\varepsilon_1| = 1$ for $M=1.6\TeV$ and $\tilde{m}_1 = 0.06\eV$,
and it is dominantly produced at low temperature $T\sim M/10$, where neglected 
higher order processes are expected correct it by $\circa{<}10\%$.

Fig.\fig{eta2}a shows how the CP asymmetry $\varepsilon_1$ needed to get the observed baryon asymmetry depends on $M$
for several values of $\tilde{m}_1$. 
Since the lepton asymmetry is dominantly generated by triplet decays at $T\sim M/20$,
the sphaleron decoupling effect becomes significant at $M\circa{>}20T_{\rm dec}\sim \hbox{few TeV}$,
preventing successful baryogenesis if $M>1.6\TeV$.

\section{Type-II see-saw: leptogenesis from a scalar triplet}  \label{II}
The two body ${\cal TT}^*$ states have total isospin $I=\{1,3,5\}$,
total spin $S=0$;
and we focus on $s$-wave annihilations so that $L=0$:
at tree level such states can only annihilate into two SM vectors, that can have isospin $I=\{1,3,5\}$.


\beq c_s =   
\underbrace{\frac{ 2g_2^4 + 3g_Y^4/2}{\pi} S_T(-(2\alpha_2+\alpha_Y) \sqrt{z})}_{I=1 \to W^a W^a,YY} +
\underbrace{\frac{5g_2^4}{2\pi} S_T((\alpha_2-\alpha_Y) \sqrt{z})}_{I=5\to W^a W^b}+
\underbrace{\frac{6g_2^2g_Y^2}{\pi}S_T(-(\alpha_2+\alpha_Y)\sqrt{z})}_{I=3\to W^a Y}
 \eeq
In the limit $S_T\to 1$ this expression for $\gamma_A$ reduces to the one in~\cite{HRS}.

Our results for the type-II see-saw are presented in fig.\fig{ann}b,\fig{eta}b,\fig{eta2}b,
that closely resemble the corresponding figures\fig{ann}a,\fig{eta}a,\fig{eta2}a already presented in the type-III section.

Fig.\fig{ann}b shows how much Sommerfeld corrections enhance the annihilation rate $\gamma_A$, thereby decreasing the efficiency. 
 Fig.\fig{eta}b shows the contour plot of the efficiency $\eta$ as function of $M$ and of $\tilde{m}_1$,
 assuming equal ${\cal T}$ branching ratios into $LL$ and into $HH$, $B_L=B_H=1/2$.
 Considering e.g.\ the point $M=10^5\GeV$ and $\tilde{m}_1  =10^{-5}\eV$, Sommerfeld corrections
reduce the efficiency from $2.9~10^{-8}$ to $2.1~10^{-8}$.

As in~\cite{HRS} we shaded in green the region of fig.\fig{eta}b where thermal leptogenesis can 
produce the observed baryon asymmetry, assuming that the CP asymmetry is generated
by the dimension-5 operator $(LH)^2$, constrained by observations of neutrino masses.

\medskip

Finally, we study the model-independent leptogenesis lower bound on $M$.
The efficiency is higher when $B_L\gg B_H$ or $B_H\gg B_L$~\cite{HRS}.
However the model-independent unitary bound on the CP asymmetry in ${\cal T}$ decays,
$|\varepsilon_L|<2\min(B_L,B_H)$~\cite{HRS}
has an opposite and stronger dependence
on $B_L,B_H$, such that the lower bound on $M$ is saturated for  $B_L=B_H=1/2$
(we are not aware of explicit models that can saturate this bound).
As in the previous section, the key point is taking into account
sphaleron decoupling, by adding the sphaleron
eq.\eq{BoltzB} to the standard equations of~\cite{HRS}.
The resulting bound $M>1.6\TeV$ is illustrated in fig.\fig{eta2}b,
which is qualitatively and quantitatively similar to fig.\fig{eta2}a.

Fig.\fig{sample}b shows the  evolution of the various asymmetries for $M=1.6\TeV$, $\tilde{m}_1 = 0.06\eV$,
and a maximal
CP asymmetry $|\varepsilon_L| = 1$.

\section{Conclusions}\label{concl}
Type-II and type-III see-saw generate neutrino masses from particles with electroweak gauge interactions:
scalar or  fermion $\SU(2)_L$ triplets.
Previous works showed that these triplets can lead to successful thermal leptogenesis~\cite{LeptogenesisTypeIII,HRS, LeptogenesisTypeIIIb} as
$n_B/n_\gamma =-0.029 \varepsilon\eta=6\cdot 10^{-10}$.
The triplet annihilation rates (that keep triplet abundances close to thermal equilibrium apparently violating the out-of-equilibrium
Sakharov  condition for baryogenesis) can be small enough to allow a large enough efficiency $\eta$.

In the present work we showed that Sommerfeld corrections, that account for long-range non-abelian electrostatic interactions,
enhance the annihilation rate, decreasing the efficiency by $\sim 30\%$.
Fig.\fig{eta}a and b show the efficiency as function of the lightest triplet mass $M$ and of the lightest triplet contribution $\tilde{m}_1$
to neutrino masses.

The efficiency steeply decreases at $M\circa{<} 3\TeV$ because, 
when at $T\sim M/20$ triplet decays produce the lepton asymmetry,
sphalerons no longer  convert it into the baryon asymmetry.
Indeed at $T\circa{<}m_h$ the electroweak symmetry starts to be broken suppressing the sphaleron interaction rate.
We find that baryogenesis via type-II or III thermal leptogenesis is only possible at $M\circa{>}1.6\TeV$.
This bound on $M$ becomes stronger if $\tilde{m}_1$ is smaller than 1 eV, or if the Higgs is heavier than its minimal value, $m_h=115\GeV$.

\medskip

Previous studies~\cite{LHCtypeII,LHCtypeIII} found that see-saw triplets lighter than about
$1 \TeV$ give detectable effects at the LHC collider.
It is therefore interesting to study if/how the absolute leptogenesis lower bound on $M$ can be evaded.
One possibility is a new source of baryogenesis at the weak scale, 
for example new physics that makes the electroweak phase transition of second order
and provides a new source of CP violation~\cite{EWB},
a scenario which is already strongly constrained but not yet excluded.
The alternative possibility is another heavier baryogenesis mechanism:
here the problem is that the baryon asymmetry gets washed out by triplets unless 
their $L$-violating interactions are slower than the expansion rate.
In the fermion triplet case, this happens if $\tilde{m}_1\ll 10^{-2}\eV$ 
(which implies significantly displaced decay vertices at LHC~\cite{LHCtypeIII})
or if the triplet is negligibly coupled to some lepton flavor.
However, neutrino oscillations need neutrino masses heavier than $10^{-2}\eV$ and with large mixings among flavors.
Scalar triplets offer one more possibility~\cite{HRS}: $L$-violating rates are suppressed if triplet decay rates
into $LL$ or $HH$ are much different, but this suppression would hold in cosmology as well as at LHC.

We obtained the bound $M>1.6\TeV$ in type-II and type-III see-saw, and we
expect a similar lower bound on $M$ in any model where the particle responsible for leptogenesis
has interactions large enough to be detectably produced at LHC.
In particular, in type-I see-saw models where baryogenesis is produced by
decays of a right handed neutrino charged under new massive vectors,
one can maybe find a situation where the new vectors are light enough
 to lead to detectable signals at LHC, and heavy enough  to suppress their contribution to $\gamma_A$~\cite{Hamb}.

We conclude discussing the bound on $M$ from the string-anthropic-multiverse point of view that 
recently received significant attention~\cite{Ant}.
Assuming that  baryogenesis is impossible  in the vast majority of string models because they predict $M\ll \langle h\rangle$,
one could argue that it is more likely to find a $M$ close to the anthropic bound
that makes baryogenesis possible.
If this possibility will be confirmed by future data,
one might even argue that the weak scale is much below the Planck scale because
it cannot be larger than $M$, which is generated
exponentially below the Planck scale by string instanton effects~\cite{Ibanez}.

\paragraph{Acknowledgements} 
I thank Antonello Polosa and Marco Bochicchio for helpful discussions about group theory.

\small



\begin{thebibliography}{nn}

\bibitem{foot}
R. Foot, H. Lew, X.-G. He and G.C. Joshi,  Z. Phys. {C44} (1989) 441;.

\bibitem{Sommerfeld}  A. Sommerfeld, Ann. Phys. 11 257 (1931).

\bibitem{LeptogenesisTypeIII}
\art[hep-ph/0312203]{T. Hambye et al.}{Nucl. Phys.}{B695}{169}{2004}.


\bibitem{HRS}
\art[hep-ph/0510008]{T. Hambye, M. Raidal, A. Strumia}{\PL}{B632}{667}{2006}.


\bibitem{MDM2}
\art[0706.4071]{M.~Cirelli, A.~Strumia and M.~Tamburini}{Nucl.\ Phys.}{B787}{152}{2007}.

\bibitem{hisano}
\art[hep-ph/0610249]{J. Hisano, S. Matsumoto, M. Nagai, O. Saito, M. Senami}{Phys. Lett.}{B646}{34}{2007}.

\bibitem{LeptogenesisTypeIIIb}
\hepart[0805.3000]{W. Fishler, R. Flauger}.

\bibitem{DMSom}
\hepart[0809.2409]{M. Cirelli, M. Kadastik, M. Raidal, A. Strumia};
\hepart[0810.0713]{N. Arkani-Hamed, D.P. Finkbeiner, T. Slatyer, N. Weiner}.

\bibitem{review}
For recent reviews see:
\hepart[hep-ph/0606054]{A.~Strumia and F.~Vissani}.
\hepart[0704.1800]{M.C. Gonzalez-Garcia, M. Maltoni}.


\bibitem{Pilaftsis}
\art[hep-ph/9607310]{M. Flanz, E.A. Paschos, U. Sarkar and J. Weiss}{\PL}{B389}{693}{1996}.
\art[hep-ph/9611425]{L. Covi, E. Roulet}{\PL}{B399}{113}{1997}.
\art[hep-ph/9707235]{A. Pilaftsis}{Phys. Rev.}{D56}{5431}{1997}.
\art[hep-ph/9803255]{E. Akhmedov, V. Rubakov, A. Smirnov}{Phys. Rev. Lett.}{81}{1359}{1998}.
\art[hep-ph/0309342]{A. Pilaftsis, T. Underwood}{Nucl. Phys.}{B692}{303}{2004}.


\bibitem{Shap}
See \art[hep-ph/0511246]{Y. Burnier, M. Laine, M. Shaposhnikov}{JHEP}{0602}{007}{2006}.

\bibitem{LEP}
The LEP Electroweak Working Group, web page
\url{http://lepewwg.web.cern.ch}.

\bibitem{BCST}
\art[hep-ph/9911315]{R. Barbieri, P. Creminelli, N. Tetradis, A. Strumia}{\NP}{B575}{61}{2000}.
See also
\art[hep-ph/0601084]{E. Nardi, Y. Nir, E. Roulet, J. Racker}{JHEP}{01}{164}{2006}.
\art[hep-ph/0605281]{A. Abada, S. Davidson, F. Josse-Michaux, M. Losada, A. Riotto}{JHEP}{09}{010}{2006}.


\bibitem{LHCtypeIII}
\art[hep-ph/0206150]{E.~Ma and D.~P.~Roy}{\NP}{B644}{290}{2002}.
\art[hep-ph/0612029]{B.~Bajc and G.~Senjanovic}{JHEP}{0708}{014}{2007}.
\hepart[hep-ph/0703080]{B.~Bajc, M. Nemevesek, G.~Senjanovic}.
\hepart[0805.1613]{R. Franceschini, T. Hambye, A. Strumia}.



\bibitem{LHCtypeII}
\art[hep-ph/0305288]{M. M\"uhlleitner, M. Spira}{\PR}{D68}{117701}{2003}.
\art[hep-ph/9606311]{K.~Huitu, J.~Maalampi, A.~Pietila and M.~Raidal}{\NP}{B487}{27}{1997}.
\hepart[0705.1495]{A.~Hektor, M.~Kadastik, M.~Muntel, M.~Raidal and L.~Rebane}.
\art[hep-ph/0304069]{E.~J.~Chun, K.~Y.~Lee and S.~C.~Park}{\PL}{B566}{142}{2003}.
\hepart[0804.1265]{W.~Chao, Z.~G.~Si, Z.~z.~Xing and S.~Zhou}.
\hepart[0805.3536]{P.~Fileviez Perez, T.~Han, G.~Y.~Huang, T.~Li and K.~Wang}.


\bibitem{EWB}
See e.g.\
\art[hep-ph/0410352]{M.~S.~Carena, A.~Megevand, M.~Quiros and C.~E.~M.~Wagner}{\NP}{B176}{179}{2005}
and ref.s therein.


\bibitem{Hamb}
\hepart[0806.0841]{J.M. Fr\`ere, T. Hambye, G. Vertongen}.


\bibitem{Ant}
O tempora, o mores!


\bibitem{Ibanez}
\art[hep-th/0609213]{L.~E.~Ibanez and A.~M.~Uranga}{JHEP}{0703}{052}{2007}.


\end{thebibliography}
\end{document}